\documentclass[twocolumn,a4paper]{article}
\usepackage[hmargin=21mm,vmargin=29.7mm]{geometry}
\usepackage{xcolor}
\usepackage{microtype}
\usepackage{newtxtext,newtxmath}
\usepackage[T1]{fontenc}
\usepackage{natbib}
\usepackage{graphicx}
\usepackage{hyperref}
\hypersetup{
    colorlinks=true,
    linkcolor=blue,
    filecolor=blue,
    urlcolor=blue,
    citecolor=blue,
    pdftitle={Finger on the pulse of asteroseismology},
    pdfauthor={Warrick H. Ball},
}

\definecolor{fg}{gray}{0.5}
\definecolor{bg}{gray}{0.9}

\title{Finger on the pulse of asteroseismology\footnote{This is a pre-copyedited, author-produced PDF of an article accepted for publication in Astronomy \& Geophysics.
The version of record \emph{Ball W., 2023, A\&G, 64, 2.30} is available online at: \url{https://dx.doi.org/10.1093/astrogeo/atad007}.}}
\author{Warrick H.~Ball}
\date{}

\begin{document}

\maketitle

\pagestyle{myheadings}
\markright{\emph{Ball W., 2023, A\&G, 64, 2.30. \href{https://dx.doi.org/10.1093/astrogeo/atad007}{doi:10.1093/astrogeo/atad007}}}

\noindent \emph{Warrick Ball highlights some recent discoveries in the
context of the past, present and future of
asteroseismology.}

\section*{Introduction}

Many stars \emph{pulsate} or \emph{oscillate}: their surfaces expand and contract periodically,
and their brightnesses and apparent line-of-sight velocities vary correspondingly.
The stars that vary the most have been known for centuries at least.  Mira ($o$ Cet), exemplar
of the Mira variables, was recorded
as a variable star at least as early as 1596 \citep{hoffleit1997}.  The first Cepheid variables,
$\eta$ Aql and $\delta$ Cep, were reported in the mid-1780s.  As astronomical instruments have become more sensitive,
astronomers have steadily identified many more classes including, to name just a few, the $\delta$ Scuti variables and $\beta$ Cephei variables during the early 20th
century and the $\gamma$ Doradus variables and slowly-pulsating B-type stars (SPBs) stars toward its end.
There is now a zoo of classes of pulsating variable stars usually named after the exemplar of that class.

It the late 1950s and early 1960s, the Sun, too, was discovered to be pulsating \citep{leighton1962,evans1962}, though with
much, much lower amplitude than observed in other pulsating stars.  Over the subsequent
decades, the nature of these pulsations, known as \emph{solar oscillations},
was understood and they became a tool with which
we could infer the deep structure of the Sun.  This is now the field we know as \emph{helioseismology}.
Some helioseismologists soon speculated on whether oscillations like the Sun's---\emph{solar-like oscillations}---
might be detected in distant Sun-like stars, which would be identified as \emph{solar-like oscillators}.
Dedicated campaigns to detect solar-like oscillations first bore fruit
when they were detected in Procyon by \citet{brown1991}.

The properties of stellar pulsations---in particular the pulsation periods---give us
important information about a star that largely complements the other
things we can observe, like its (average) brightness or temperature.  At the simplest extreme,
a single mode tells us about how long it takes for pressure perturbations to be restored
to equilibrium, which is closely related to a star's average density.  Even such a simple
connection leads to tight relationships between the pulsation periods and intrinsic luminosities
of certain kinds of pulsating star.  Henrietta Leavitt famously determined such a period--luminosity
relation for the classical Cepheids, with which Edwin Hubble showed that some nebulae,
like what we now know is the Andromeda Galaxy, could not be part of the Milky Way.

At the other extreme, when multiple pulsations are present, each
distinct period offers slightly different information about a star's
interior, so the observation of many modes allows us to ``see'' inside
the star.  We have measured many hundreds of pulsation periods in the Sun and combined
them to map the density, speed of sound and rotation rate in most of its interior.
These constraints demonstrated
that the apparent lack of neutrinos emanating from the Sun was a problem for
particle physics, not stellar physics.  The tension was resolved by the discovery
of neutrino oscillations, for which the Nobel Prize in Physics was awarded in 2015.

By the end of the 20th century, we had clearly already learned a great deal from
stellar pulsations.  The study of stellar pulsations, now known as \emph{asteroseismology},
was about to undergo its first space-based revolution.

\section*{The long journey to space-based photometry}

When observing periodic variations of any sort, there are three qualities to aim for.
First, one should observe with enough sensitivity to detect the periodic variations, either
by using a sensitive detector or observing for a long time.  Secondly, one should observe
for at least as long as the periodic variation being studied, and preferably longer, to
be able to distinguish between variations with similar periods.  Longer observations also
increase the overall sensitivity, in essence because most noise (any variability we aren't interested in)
is spread more evenly across all periods.  Finally, it helps to observe
as close to continuously as possible.  Interruptions lead to ambiguity about the precise
period of variation.

For astronomical observations, these considerations raise a few issues.  
The first is
the duration of observations.  To distinguish between the closely-spaced periods in some
pulsating stars, one needs observations that last many months.  That is in principle possible
except that Telescope Allocation Committees aren't always happy to commit so much time to a
single star when other astronomers are yearning to look briefly at many other things.
Second, one (typically) cannot observe continuously from a single site for more than about 12 hours.
It is possible to combine observations from multiple telescopes or observe certain stars
from near the North or South Pole (though only the South Pole is at all practical)
but is difficult to co-ordinate and still requires a lot of telescope time.
Finally, some classes of pulsating star only pulsate at very low amplitudes,
which limits observations either to very bright targets or very large telescopes.

This is not to say asteroseismology is impossible from the ground.
Far from it!  Clearly, every discovery I've mentioned so far was made
using ground-based observations and even now many ground-based projects
are produced outstanding data.  In 1986, astronomers from the University of Texas
set up a project that became the \emph{Whole Earth Telescope} (WET), in which
various telescopes around the world co-ordinated their observations to monitor
pulsations in white dwarfs \citep{nather1990}.  A similar principle is being
pursued by the Stellar Observations Network Group, which aims to continuously monitor solar-like
oscillations in bright stars \citep{song2019}.  Asteroseismology has also been
possible using long-term projects that have been monitoring for other transient
events.  The \emph{Optical Gravitational Lensing Experiment} (\emph{OGLE}),
for example, was largely intended to monitor many stars for gravitational microlensing,
where a distant star briefly appears brighter because an object between
focusses its light towards us.  OGLE's data was used in the discovery of what
we now call the \emph{blue large-amplitude pulsators} \citep[\emph{BLAP}s,][]{blaps2017},
which are intrinsically rare and probably the precursors of other small pulsating stars.
The \emph{All Sky Automated Survey for SuperNovae} (\emph{ASAS-SN}) and \emph{Wide-Angle
Search for Planets} (\emph{WASP}) have also published extensive catalogues of
variable stars.  These ground-based observations will always have their place
because they can keep operating for longer than space-based missions at much
lower cost.

Notwithstanding these successes, 
the community knew that observations from space would overcome many of the obstacles
of ground-based observations, particularly for solar-like oscillations.
By the early 1990s, teams in the United States and Europe were proposing
roughly 1-m class space telescopes that would record brightness variations in a large
number of stars simultaneously.  The larger European proposal became known
as \emph{Eddington} and was abandoned in 2003,
though in many ways laid the foundation of what is now the European Space Agency's (ESA's)
\emph{Planetary Transits and Oscillations (PLATO)} mission.
The American proposal, which was really a planet-finding mission to be used
for asteroseismology on the side, eventually became \emph{Kepler} and launched in 2009.

Even before \emph{Kepler} launched, however, other smaller missions played an important
role.  In particular, the French-led \emph{COROT} satellite detected solar-like oscillations
in a about a dozen main-sequence stars and thousands of red giants.
Among other things, its data showed without a doubt that
not only did red giants show solar-like oscillations, but their pulsations were rich
with detail \citep{deridder2009}.  What's more, this detail can be used to relatively easily
infer the red giants' masses, and thus their ages, which is critical
for understanding how the Milky Way formed.
Many of the principal techniques with which we now study solar-like oscillations
were refined on the basis of \emph{COROT}'s observations.

\section*{The first space-based revolution}

Then came \emph{Kepler}.  There is absolutely no doubt in the asteroseismology community
that \emph{Kepler} revolutionised the field.  The data that it
produced during its original mission, as well as its second life as K2, remains unparalleled
in duration and quality and is still being
explored.  I shall only highlight a few results from its nominal mission.
Naturally, most are for pulsations whose analysis benefits most from \emph{Kepler}'s strengths.

One group of pulsating stars whose study has benefitted from the long time series
are the red giants \citep[see ][for a review]{hekker2017}.
Early results from \emph{Kepler} demonstrated that it was
possible to distinguish the speed with which a red giant's core and envelope
rotated \citep[e.g.][]{beck2012}.  Surprisingly, the cores turned out to rotate much more slowly
than cutting-edge models predicted \citep{eggenberger2012, marques2013}.
This is still not fully explained and
scientists are still proposing and testing different physical processes
by which the cores might transfer some of their rotation to the envelopes.

Astronomers also noticed that in some red giants, some of the
pulsations are much weaker than in other red giants, even if they are
otherwise very similar \citep{mosser2012}.  \citet{fuller2015} proposed that strong magnetic
fields in the cores of the red giants could dampen selected pulsations
in a certain range of periods.  \citet{mosser2017} contested this
interpretation, on the grounds that the pulsations are present, though weak,
and the proposed mechanism would suppress them completely.
Recently, \citet{ligang2022} reported direct evidence of strong core magnetic fields,
again from asteroseismology, but there is still no consensus
on what causes the selective dampening.

Another group of stars that benefit from \emph{Kepler}'s long
observations are the $\gamma$ Doradus variables.  These stars pulsate
in a large number of modes that fall into groups with very similar
periods.  What's more, these pulsation periods are often frustratingly
close to one cycle per day, so ground-based observations, even if
taken for long enough, are easily confounded by only being able to
observe with the same frequency as the pulsations themselves.
\emph{Kepler}'s long, almost uninterrupted observations have thus led to an enormous leap in our understanding
of these stars.  Among other things, \citet{ouazzani2017} demonstrated
how one can measure the rotation rate inside these stars
using the patterns among the many observed modes.  \citet{ouazzani2020}
further showed that there are also cases where the visible pulsations
in the outer layers couple to a different class of pulsations near
the centre.

\emph{Kepler} also demonstrated the synergy between asteroseismology and
the search for (and study of) exoplanets.  Consider for a moment what it
takes to detect and study the small drop in brightness as an exoplanet
passes between us and its host star, known as a \emph{transit}.  First,
one wants to watch the star for as long and as continuously as possible,
to avoid missing a transit.  Second, one wants to use very sensitive
detectors, since an Earth-like planet would only occlude about one
ten-thousandth of a Sun-like star's light.  These requirements should
sound familiar: they're precisely what we seek in asteroseismic observations.
Thus, pulsations were often detected in the same \emph{Kepler} data
used to detect planets.  This allows a better characterisation of the star,
which sets the absolute scales for the exoplanets.  It also led to
some intriguing results, like the misalignment of \emph{Kepler}-56
and its exoplanets \citep{huber2013}.
The synergy between asteroseismology and exoplanetology is at the core
of the upcoming \emph{PLATO} mission.

After the second of \emph{Kepler}'s four
reaction wheels failed, the spacecraft could no longer point accurately
enough for the primary mission, which thus ended.
The satellite's detectors were still operating, however, and by ingeniously
balancing the satellite (unstably) against the solar wind, \emph{Kepler}
continued to observe under a new mission: \emph{K2}.  All the stars to be observed
were now selected by proposal and a remarkable variety of objects were
observed, including Neptune, in whose reflected light \citet{gaulme2016neptune} detected
the Sun's oscillations.
One asteroseismic highlight was the study of variability among the seven brightest
stars of the Pleiades cluster---all B-type giants or supergiants---by
\citet{white2017pleiades}, which includes \emph{Maia}.
\citet{struve1955} claimed to detect variability in \emph{Maia} and proposed
a new class of \emph{Maia} variables whose existence quickly became contentious
and is still debated.  Whatever the Maia variables are, and
whether or not the class even exists, \citet{white2017pleiades} argued
convincingly that ``Maia is variable, but Maia is not a Maia variable''.

Though \emph{Kepler}'s observations were of unparalleled quality,
it only observed stars in a relatively small region of the sky---about
115 square degrees---, prioritised cool stars around which
planets were more likely to be found and only observed a few hundred
main-sequence stars often enough to potentially analyse
solar-like oscillations.  \emph{K2}'s community-driven target selection led to a greater range
of science than the primary mission but its observations were still
limited to relatively small fields, all along the ecliptic.
\emph{Kepler}'s observations were also limited to relatively faint
stars for which complementary observations were difficult.  With the
new insight that exoplanets are common, the next step would be to
survey brighter stars over a wider area, which was precisely the idea behind
NASA's \emph{Transiting Exoplanet Survey Satellite} (\emph{TESS}), led
by George Ricker.

\section*{A second revolution}

\emph{TESS} is a distinctly different mission from \emph{Kepler}.  It
is has a smaller collecting area---just four cameras with 10 cm
apertures---that covers a 96 by 24 degree segment of sky
for about 27 days at a time.  These fields of view overlap to various extents,
depending on the satellite's planned observations, with a maximum continuous
overlap of about a year achieved for stars near the ecliptic poles.

The time series are neither as long nor as precise as \emph{Kepler}'s.
What \emph{TESS} lacks in quality, however, is more than made up for
by the quantity of data.  The entire field of view is recorded at intervals
that have shortened as the mission has continued.  During the original two-year mission,
these full-frame images (FFIs) were taken every 30 minutes and covered about 85 per cent
of the sky.  In the first mission extension,
the imaging interval decreased to 10 minutes.  In the second extension, it will decrease to just 3 minutes
and 20 seconds!  In addition, some stars are selected by open proposal to be observed
every two minutes and an additional observing mode to take data every twenty seconds
was added in the first extended mission.  This has provided a truly staggering amount
of data, especially to those fields of asteroseismology---particularly of hot stars---that
were neglected by \emph{Kepler}'s planet-hunting mission.  I think it's safe to say
that we have not kept up with the data: there's just too much!
But we've found plenty of interesting things
where we have had time to look.

Let's begin with the solar-like oscillators.  \emph{TESS}'s
observations are, expectedly, not as sensitive as \emph{Kepler}'s, so
our analyses of main-sequence and subgiant stars have been limited to \emph{very} bright stars
($V\lesssim6$, naked eye in ideal conditions).  Instead of being
limited to the hundred or so stars that happened to be selected for
observations, however, we can now choose stars that are
otherwise interesting and search their \emph{TESS} data for solar-like
oscillations.

So far, this has mostly been limited to individual systems,
including stars that were already known to host planets, e.g.
$\lambda^2$ Fornacis \citep{nielsen2020}, HD~38529 \citep{ball2020}, HD 212771 and HD 203949 \citep{campante2019}.
These complement the measurement of solar-like oscillations in two systems---HD 221416 \citep{huber2019} and HD 19916
\citep{addison2021}---whose planets were themselves detected in TESS data.
\citet{hatt2023} very recently searched for solar-like oscillations in the TESS
data of over 250\,000 stars, finding solar-like oscillations in 4\,177 of them.
The catalogue includes 28 known planet hosts, so more individual studies like
this are possible.

Exoplanets aside, there are also new analyses of systems that are interesting for stellar physicists,
of which I shall mention three examples.
The binary system 12 Boo comprises two stars of which the more
massive component is only a few percent heavier but about 60 per cent brighter
than its companion.  This can only mean that the stars are in a very particular
phase of evolution, in which the pulsations would tell us a great deal about
how the stars' cores have evolved.  Though \citet{ball2022} were unable to
measure the individual pulsation periods, the broader characteristics of
the observations have led to more precise parameters for the system.

\citet{chontos2021} studied the solar-like oscillations of the solar
analogue $\alpha$ Men, which is orbited by a low-mass red dwarf.
Low-mass stars evolve very slowly, which makes it notoriously
difficult to determine how old they are: they look much the same
after one billion years as they do after ten.  By measuring the brighter
star's age through its oscillations, Chontos et al.~also precisely constrained the age of
the low-mass companion.

Finally, \citet{chaplin2020} studied the solar-like oscillations of
the metal-poor subgiant $\nu$ Indi, which was observed by TESS near
the beginning of the mission.  
This star formed early in the Milky Way's life and its motion through
the galaxy has been affected by the Milky Way's interactions with
other galaxies as it has grown over many billions of years.
In particular, $\nu$ Indi's orbit was affected by the Milky Way's
last major merger with another dwarf galaxy, known as \emph{Gaia Enceladus}.
Because the merger must have happened after $\nu$ Indi formed,
the star's age of about 11 billion years---inferred using asteroseismology---places
an upper limit on how long ago the merger could have happened.

\begin{figure}
  \includegraphics[width=\columnwidth]{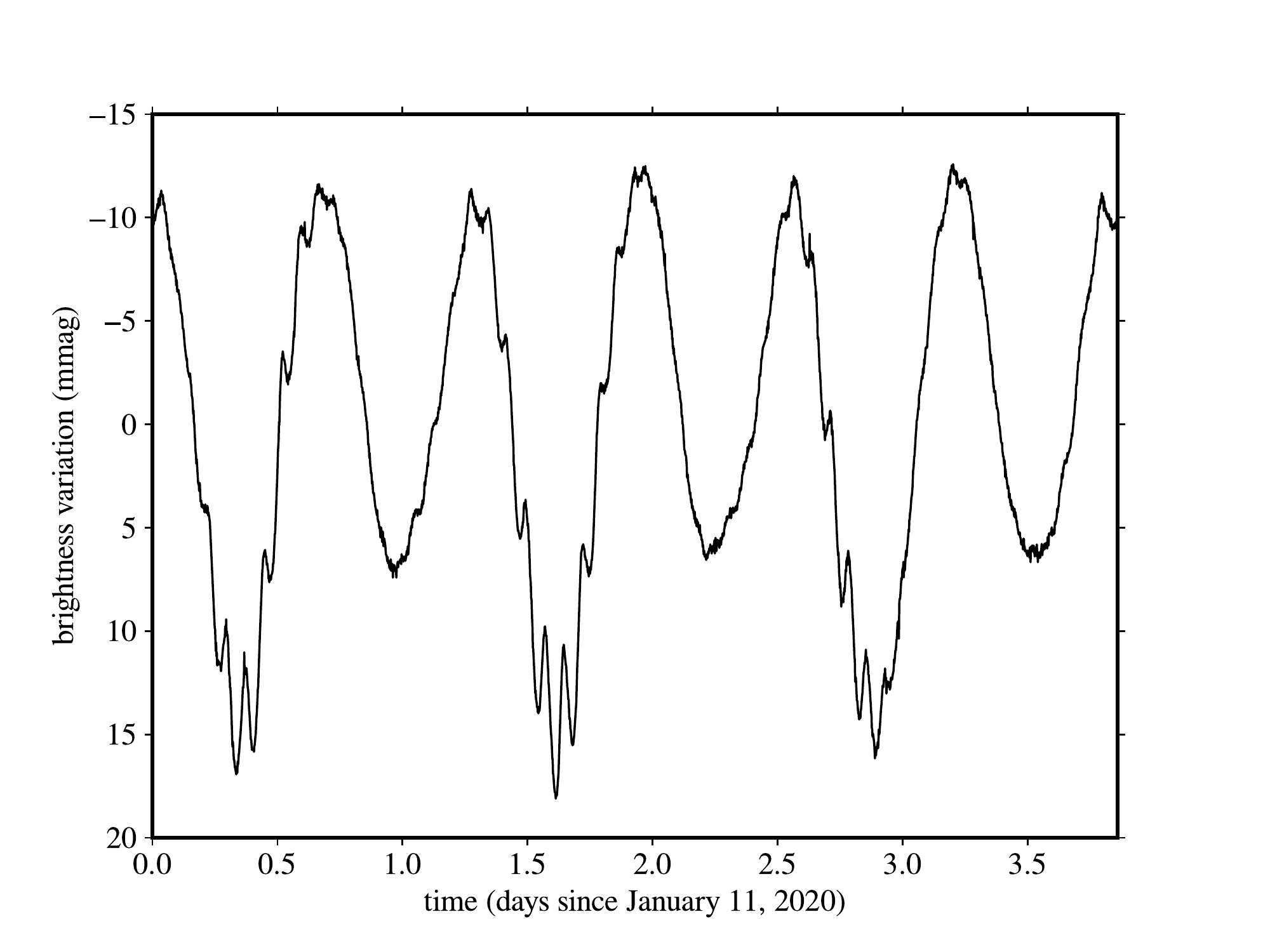}
  \caption{Brightness variations over a few days in the single-sided
    pulsator CO Cam.  The long-period variation is three orbital
    cycles in which the brightness changes because the brighter star
    is distorted into a slight teardrop shape by its companion.  The
    short-period variations are its pulsations, which, unexpectedly,
    have higher amplitudes in the troughs of the orbital variation.}
  \label{f:co_cam}
\end{figure}

These results are interesting examples of how TESS has allowed us
to take the lessons learned from \emph{Kepler} and apply them to
other solar-like oscillators.  But \emph{TESS}'s revolutionary results
in asteroseismology have been in other types of pulsating stars.
As a first example, consider what we now call \emph{tidally trapped}
pulsations.  \citet{handler2020} discovered pulsations in the A-type
binary star HD 74423.  The apparent brightness of the primary varies
because it's distorted into a teardrop shape by the gravity of its companion.
In addition, the pulsations are modulated during the orbit such that they
must be somehow confined to the side of the primary pointing either
toward or away from the companion.  More of these tidally trapped
pulsators have been detected in TESS's data \citep[e.g. in CO Cam,][Fig.~\ref{f:co_cam}]{kurtz2020}
and they offer an opportunity
to study the interaction between pulsations and tidal forces, the latter
of which influence how binary stars evolve.

A further revolutionary discovery among the $\delta$ Scuti variables
has been a group of stars that pulsate at a large number of relatively
short periods.  To understand the value of these results, let me first digress
on the importance of \emph{mode identification}.  To compare observed pulsation
periods to predictions from stellar models, we need to know
which mode we're actually seeing.  In certain regimes, this is quite easy.
Solar-like oscillations in main-sequence stars, for example, follow certain
formulae called \emph{asymptotic relations} that
allow us to tell which mode is which.  For other
kinds of pulsating star, including many $\delta$ Scuti variables, this is much more
difficult.  The asymptotic relations aren't valid and there's no guarantee
that some modes simply aren't excited to measurable levels.

\begin{figure}
  \includegraphics[width=\columnwidth]{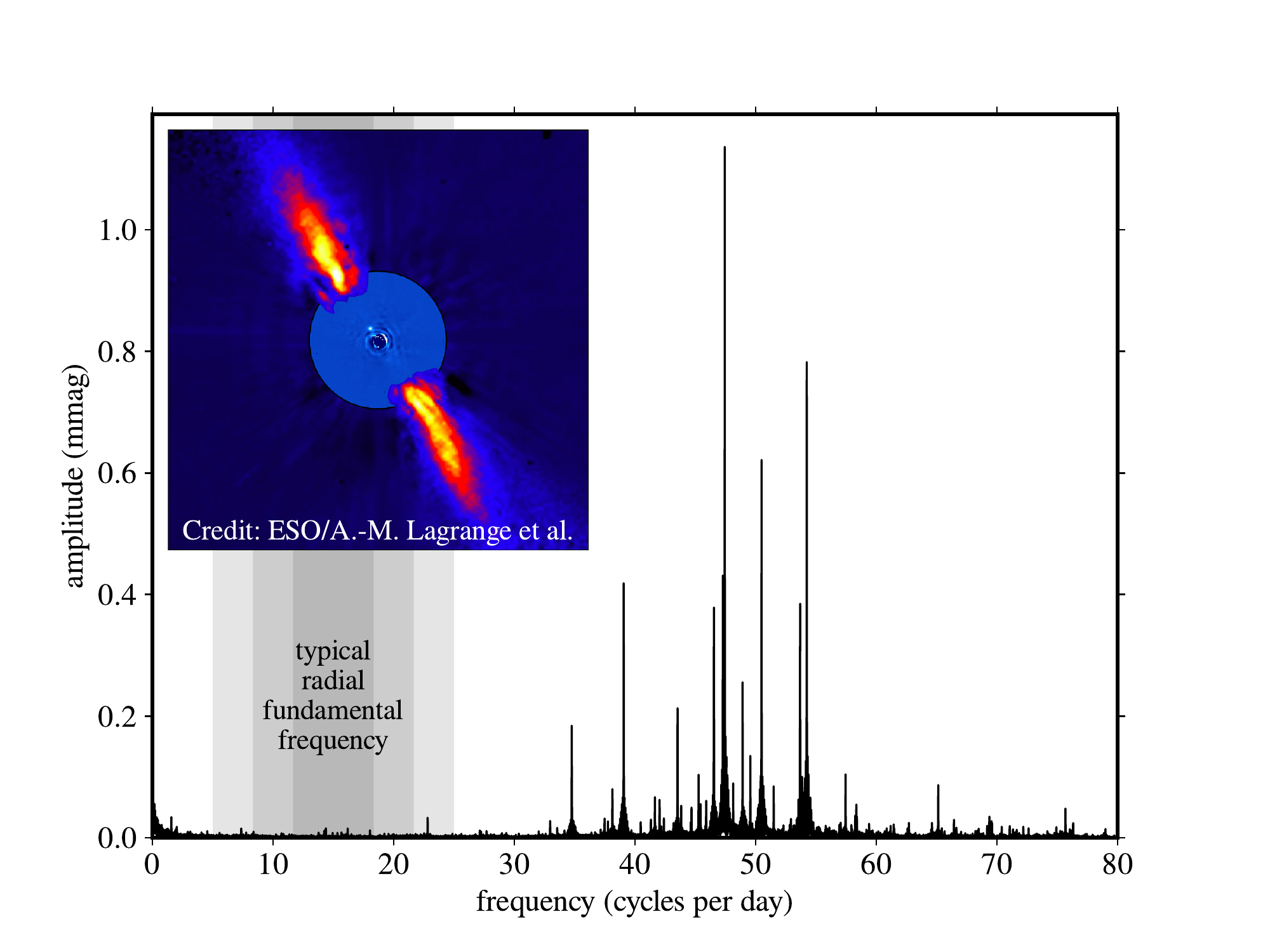}
  \caption{A periodogram of the young planet host $\beta$ Pictoris, showing
    multiple pulsations at frequencies higher than usually found in $\delta$
    Scuti variables.  The inset shows a composite infrared image of
    the system, which includes two planets and a debris disk.}
  \label{f:bet_pic}
\end{figure}

\citet{bedding2020} serendipitously discovered a group of $\delta$
Scuti variables in which a large number of modes are excited at
shorter periods than the typical fundamental pulsation.  These
pulsations \emph{do} satisfy a sort of asymptotic relation, which has
opened up the opportunity to precisely characterise the pulsating
stars, because now we can tell which modes are which.  It's also a
happy coincidence that these are young stars and some (e.g.
$\beta$ Pic, Fig.~\ref{f:bet_pic}) have features that suggest that planets are forming
or recently formed.  \citet{murphy2021} analysed one of these
short-period $\delta$ Scuti variables to very precisely constrain the
age of the stellar association (a kind of loose star cluster) of which
it is part.  With hindsight, astronomers are also revisiting
\emph{Kepler}'s observations \citep{murphy2022}.

As a final example, \citet{dorn-wallenstein2019} searched for rapid variability in the yellow supergiants in the
Large Magellanic Cloud (LMC).  The LMC is near the Southern Ecliptic Pole and
TESS observed it almost continuously for the first year of its observations.
The team quickly found a handful of yellow supergiants that appear to be pulsating
with periods shorter than a day.  Though they initially dubbed these stars the
\emph{fast yellow pulsating supergiants}, we still don't know quite what these stars are.
They might be cooler versions of the $\alpha$ Cygni variables or something else entirely.
Whatever the nature of their pulsations, they may well be red supergiants that lost
a significant fraction of their envelope and are about to (in astronomical terms) become
supernovae.

\section*{The revolutions yet to come?}

TESS has already been observing the sky for over four years and has
funding to do so until late 2024.  Its novel orbit, which resonates
with the Moon, should be stable for decades and require little fuel to
maintain.  Its exciting to consider the potential for over a \emph{decade}
of TESS observations but the satellite still has some limitations.

One of TESS's limitations is that, particularly for the study of solar-like oscillations,
which we hope to find in cool, main-sequence planet hosts, TESS is not
particularly sensitive.  \emph{Kepler} was better but had a limited field
of view and ultimately only observed a few hundred solar-like oscillators
very precisely.  Two upcoming missions plan to fill this gap.
The first is \emph{Earth 2.0} \citep[ET;][]{et},
which is being developed in China.  ET is in many ways a successor to
\emph{Kepler} and intends to monitor a region of sky about four times
larger than (and selected to include) \emph{Kepler}'s field-of-view
with the aim of detecting ``Earth twins'': Earth-like planets in
Earth-like orbits around Sun-like stars.  By designing the mission to
mitigate \emph{Kepler}'s noise sources, the aim is detect perhaps
ten times as many planets but the data will again be valuable for
the study of pulsating stars.

The second mission is the European Space Agency's (ESA's)
Planetary Transits and Oscillations mission \citep[PLATO;][]{plato}.  The synergy
between solar-like oscillations and exoplanet transits is at the core
of PLATO's mission, so it has been designed to detect solar-like
oscillations in thousands (and perhaps tens of thousands) of
solar-like oscillators.  PLATO will have a larger overall field of
view than ET (about twenty times larger than \emph{Kepler}) but its
sensitivity varies across that field of view and it will concentrate
on brighter stars.  Both ET and PLATO are currently expected to launch
by the end of 2026.

Another limitation of TESS (and one shared by PLATO and ET, to different extents)
is the large size of its
pixels.  Many modern astronomical detectors have pixels, much like ordinary digital cameras.
Each pixel in TESS's detectors spans quite a large
patch of sky---about 21 arcseconds across---which makes it difficult to work out
which star's light is varying when there are many stars in a small region
of the sky.  This ``blending'' often happens in star clusters, particularly the old globular clusters
that surround the Milky Way.  Understanding the stars in these environments
would offer many insights and a suitably small pixel scale is the basis of a proposed mission
dubbed \emph{High-precision AsteroseismologY in DeNse stellar fields} \citep[HAYDN;][]{haydn}.
HAYDN was very recently selected as one of five candidates from which
ESA will select its next medium-class mission, which would launch some time
in the late 2030s.

TESS should operate for many more years, two new missions are being prepared for launch in
a few years' time and another mission might launch before 2040.
It's clear that space-based photometry has already revolutionised the study of stellar pulsations.
It is difficult to believe that there are not more revolutions waiting in the data.

\clearpage

\section*{Boxes}

\subsection*{What is the nature of stellar pulsations?}

  Many stars, including the Sun, rotate slowly enough that they
  can be treated as perfectly spherical.  It's worth keeping in
  mind that some stars (e.g.~Altair) rotate so fast they are
  measurably oblate.  If we also assume the stars are alone, stable over
  short timescales (which for stars might mean millenia) and not
  magnetised, we can compute useful models of how stars change over
  long timescales.
  Given one of these models, we can then ask what happens over short
  timescales if the star is somehow perturbed.  As long as any subsequent pulsations
  are fast, we can further assume that the pulsation is an entirely mechanical
  thing: the different layers move back and forth without affecting
  how much heat is exchanged between them.

  All these assumptions lead to a set of equations that we can solve
  to determine how a star will vibrate if it is perturbed.  We'll
  leave to another box the question of how a star is perturbed in the
  first place.  If we presume the pulsations happen, we find that
  their dependence on the depth inside the star separates them into
  two families, distinguished by which force pushes the perturbed star
  back towards equilibrium.  In both cases, there are certain depths within
  the star where there is not motion during the pulsation: these are the \emph{nodes}
  and their number is the \emph{radial order} $n$ of a mode.

  First, there are \emph{pressure modes} or \emph{acoustic modes}.
  Here, as the name implies, the perturbation is restored by the
  pressure, just like the perturbations in the air around you that you
  perceive as sound.  These are the modes that we observe in the Sun,
  main-sequence solar-like oscillators, $\delta$ Scuti variables,
  $\beta$ Cephei variables, RR Lyrae variables and Cepheid variables,
  among others.

  Second, there are \emph{gravity modes}.  Here, the perturbations are
  instead restored by buoyancy of denser layers trying to rise into
  more rarefied layers.  These are observed, for example, in the
  $\gamma$ Doradus variables and slowly-pulsating B-type stars.

  There are also stars, particularly red giants, in which some
  pulsations are simultaneously like pressure modes near the surface
  and gravity modes nearer the centre.  We call these
  \emph{mixed modes}.  The coupling between the two varieties of modes
  changes the frequencies from those of pure pressure modes, revealing
  properties of the core.  The mixed modes are much of the reason that
  \emph{Kepler}'s observations of red giants have been so powerful.

  All the observed modes can be further classified by how they vary
  across the surface of the star.  In this case, there are lines
  across the star's surface (really, planes through the whole star) that don't move during the pulsation.
  Their total number is the \emph{angular degree} $\ell$ and the number
  that pass through the star's poles is the \emph{azimuthal order} $m$.
  \emph{Radial} modes have $\ell=0$, so every layer of the star expands
  or contracts in unison.  In the case that all the layers move together
  as a function of depth, too, we have the \emph{radial fundamental} mode,
  which is the most common pulsation mode among the classical pulsators.
  Modes in which the star pulsates differently at different latitudes and
  longitudes are \emph{non-radial} modes.
  
  The spherical harmonics turn up whenever we solve an equation for
  waves in spherically symmetric system.  In quantum mechanics,
  this situation arises in the atomic orbitals too so the wavefunctions
  of electrons are also described by spherical harmonics.

  I'll finish by returning to the original assumption: stars don't
  rotate.  In truth they do and when the rotation is slow, this gently
  causes the modes of the same angular degree $\ell$ but different
  azimuthal order $m$ to pulsate at slightly different frequencies.
  If we can measure these slight differences, we can learn about how a
  star rotates beneath the surface, and, for the Sun, we have even
  done so as a function of both depth and latitude!

  For fast rotating stars, the equations become dramatically
  more complicated.  But in the last decade, we have discerned
  pulsations in many stars, including the Sun itself, that exist
  \emph{because} of rotation.  These are the \emph{inertial} modes,
  restored by the Coriolis force,
  and their rich insights are on offer to all those who can afford to
  toil through their (relative) theoretical complexity.

\begin{figure}
  \includegraphics[width=\columnwidth]{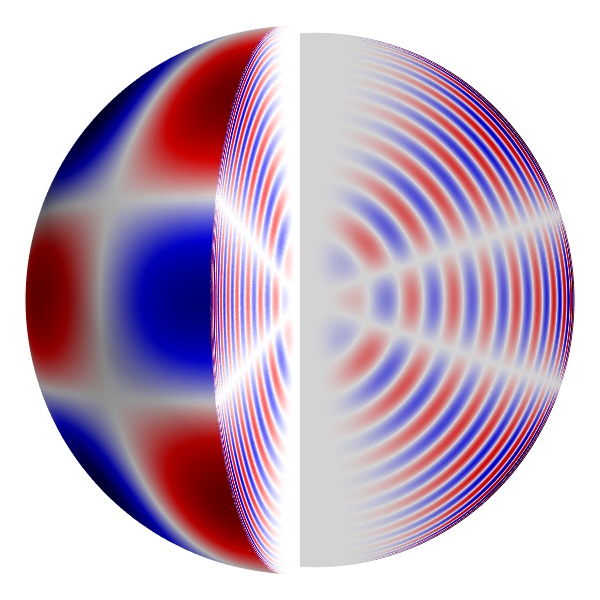}
  \caption{The eigenfunction of a pressure mode in a standard model of
    the Sun, with angular degree $\ell=6$, azimuthal order $m=4$ and a
    frequency of about $3\,\mathrm{mHz}$.  Alternating red and blue regions would
    be moving in opposite directions at a given phase
    of the pulsation.}
\end{figure}

\subsection*{What drives stellar pulsations?}

  For a star to pulsate, something must \emph{drive} the pulsations.
  Two reasonably well-understood mechanisms cover most
  pulsating stars.  First, there is the \emph{$\kappa$-mechanism}, so
  called because it is related to how a star's \emph{opacity}---denoted
  by $\kappa$---varies with temperature.  Imagine a layer beneath the surface
  of a star where the opacity happens to increase as the temperature decreases.
  If that layer absorbs a bit more heat than the long-term average,
  it will expand slightly and cool.  But by cooling, it becomes more opaque
  and absorbs \emph{more} heat.  This process runs away until the opacity
  starts to decrease with decreasing temperature again, at which point
  the excess heat is able to escape.  The layer contracts, the temperature
  rises again, and the cycle repeats.  \citet{eddington1917} observed that
  the relevant layer of a star acts like a heat engine and the mechanism
  is still occasionally referred to as the \emph{Eddington valve}.

  The cause of the opacity \emph{bumps} that destabilise stars turns
  out the be the changing ionisation states of some of its atoms.  The
  Cepheid, RR Lyrae and $\delta$ Scuti variables are destabilised by
  helium going back and forth between its first and second ionisation
  states.  The $\beta$ Cephei variables and slowly-pulsating B-type
  stars are destabilised by a feature associated with a number of metals,
  including iron, around a few hundred thousand kelvin.  The same
  mechanism is thought to destabilise the variable subdwarf B stars.

  The other broad driving mechanism is near-surface convection,
  which is responsible for all solar-like oscillations, be they
  in main sequence stars, subgiants or red giants.  In
  stars with surfaces cooler than about 7\,000 kelvin, radiation alone
  is unable to transport all of the heat from nuclear fusion into
  space.  The plasma becomes unstable to convection and large,
  turbulent flows develop.  These flows become very strong---they can
  travel at a good part of the speed of sound near the surface---and their turbulent
  motions perturb the whole star, like a continuously rolling drum.
  The convective motions also dampen the oscillations but the star nevertheless
  resonates measurably at certain frequencies.
  It's fair to say that this mechanism is not fully understood
  though the principles are well-established.

  These two mechanism are not the whole story.  There are several
  classes of pulsator for which there is no consensus on what drives
  their pulsations.  One class is the $\gamma$ Doradus variables,
  whose luminosities and temperatures are very similar to the $\delta$ Scuti variables.
  Another class are the Mira variables, whose pulsations might be driven by
  the changing ionisation state of hydrogen and helium, by near-surface
  convection or by some combination of the two.

\subsection*{Why are there so many classes of pulsating stars?}

  Each class of pulsating star can be thought of as an intersection of four
  properties.  First, there are only certain conditions under which stellar pulsations
  can be driven, usually by the $\kappa$-mechanism or near-surface convection.
  Second, there is the variety of pulsation mode that is driven, where we usually
  distinguish pressure modes, gravity modes and mixed modes.
  Third, stars are born with a distinct distribution of properties, which is related
  to how the Milky Way evolved and how stars form from collapsing clouds of gas.
  Finally, as they evolve, stars only ever obtain certain combinations of properties
  and those properties change at varying rates.

  Take, as an example, the \emph{classical instability strip}.  This is a region
  of temperature--luminosity space in which stars are unstable to the $\kappa$-mechanism
  because of the second ionisation of helium.  The instability strip itself spans
  luminosities from 10 to 100\,000 times the Sun's luminosity and surface temperatures
  from about 8\,000 kelvin at low luminosity down to about 5\,000 kelvin at high luminosity.
  The strip is not completely full of stars because they evolve at different rates
  across the instability strip at different times in their lives.  When a star about twice
  as massive as the Sun is on the main sequence, it might spend much of its life in the
  instability strip as a $\delta$ Scuti pulsator.  When a star five times as massive as the Sun
  has begun fusing helium into carbon and oxygen in its core, it might evolve across
  the instability strip as a classical Cepheid variable.
  If the $\delta$ Scuti variable is relatively lacking in heavy elements,
  we might identify it as an SX Phoenicis variable instead, even though much
  of the underlying physics is the same as the $\delta$ Scuti variable.
  All three classes just mentioned are likely to pulsate in the radial
  fundamental mode, though many $\delta$ Scuti variables also pulsate in higher order modes.

  As another example, consider the $\beta$ Cephei variables and slowly-pulsating B-type stars.
  Both classes are generally B-type stars near the main-sequence, in which
  oscillations are driven by an opacity feature related to heavy elements, including iron.
  The two classes are distinct mainly because the $\beta$ Cephei variables pulsate in
  low-order pressure modes (like the $\delta$ Scuti variables) whereas the slowly-pulsating B-type
  stars' pulsations are gravity modes (like the $\gamma$ Doradus variables).
  That the B-type variables are rarer than the lower-mass variables is largely
  unrelated to pulsation.  More massive stars simply form less often than less massive stars.

  This framework, though useful to understand the classification, does not
  offer a complete understanding of stellar pulsations.  A critical point is
  that we find stars in instability strips that are \emph{not} pulsating
  and it is unclear why.
  The waters are also muddied somewhat by stars that have probably
  interacted with companion stars.  As our observations grow ever more numerous,
  so too we discover new, intrinsically rare classes of pulsators
  that we must somehow fit into our schemes above.
  Like much of astronomy, now as in the past, new
  classes of pulsating stars tend to be invented before we fully
  understand what we are actually looking at.

\begin{figure*}
  \includegraphics[width=\textwidth]{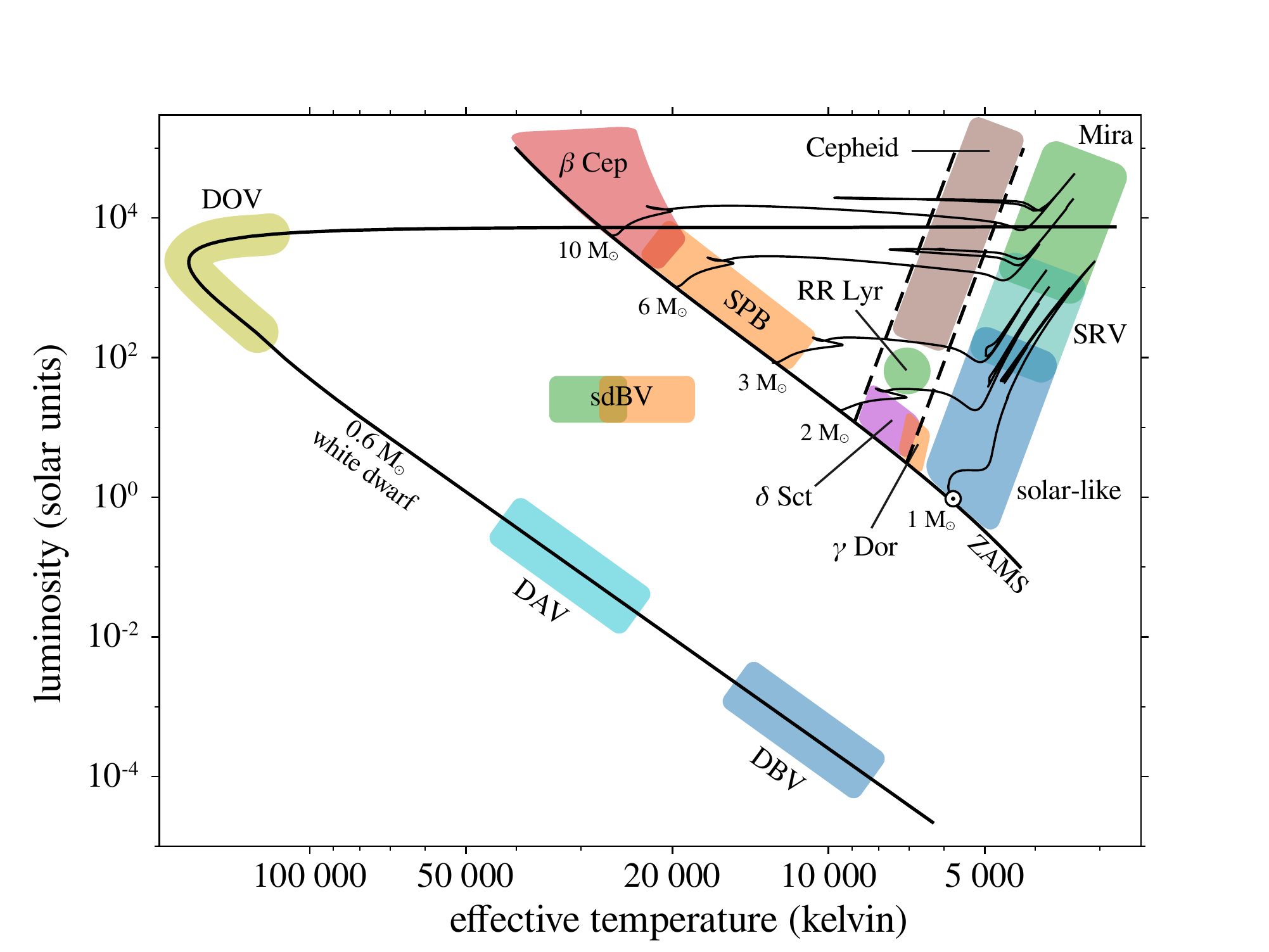}
  \caption{A Hertzsprung--Russell diagram showing typical surface
    temperatures and luminosities of many (but not all!) types of pulsating star.
    The different colours are only to help distinguish the different
    classes.  Thin solid lines show theoretical evolutionary sequences
    for stars of the indicated masses.  The thick solid lines show the
    zero-age main sequence (ZAMS, where hydrogen fusion begins) and the
    approximate evolution of a $0.6\,M_\odot$ white dwarf.}
\end{figure*}

\clearpage
\footnotesize
  
\section*{Author}

Warrick H.~Ball is a Senior Research Software Engineer in the Advanced
Research Computing team at the University of Birmingham.  Until
recently, he was a Postdoctoral Fellow in the School of Physics \&
Astronomy, where he tried to make model stars pulsate at the same
frequencies as real ones.
  
\bibliographystyle{../mn2e}
\bibliography{../master}

\end{document}